\documentclass[prl,twocolumn, superscriptaddress]{revtex4}

\usepackage{graphicx}
\usepackage{dcolumn}
\usepackage{bm}
\usepackage{braket}
\usepackage{color}

\begin{document}

\title{Surface valence transition in SmS by alkali metal adsorption}

\author{Takuto Nakamura}
\email{nakamura.takuto.fbs@osaka-u.ac.jp}
\affiliation{Graduate School of Frontier Biosciences, Osaka University, Suita 565-0871, Japan}
\affiliation{Department of Physics, Graduate School of Science, Osaka University, Toyonaka 560-0043, Japan}

\author{Toru Nakaya}
\affiliation{Department of Physics, Graduate School of Science, Osaka University, Toyonaka 560-0043, Japan}

\author{Yoshiyuki Ohtsubo}
\affiliation{Institute for Advances Synchrotron Light Source, National Institutes for Quantum Science and Technology, Sendai 980-8579, Japan}
\affiliation{Graduate School of Frontier Biosciences, Osaka University, Suita 565-0871, Japan}
\affiliation{Department of Physics, Graduate School of Science, Osaka University, Toyonaka 560-0043, Japan}

\author{Hiroki Sugihara}
\affiliation{Department of Physics, Graduate School of Science, Osaka University, Toyonaka 560-0043, Japan}

\author{Kiyohisa Tanaka}
\affiliation{Institute for Molecular Science, Okazaki 444-8585, Japan}

\author{Ryu Yukawa}
\affiliation{Graduate School of Engineering, Osaka University, Suita 565-0871, Japan}

\author{Miho Kitamura}
\affiliation{Photon Factory, Institute of Materials Structure Science, High Energy Accelerator Research Organization (KEK), 1-1 Oho, Tsukuba 305-0801, Japan}

\author{Hiroshi Kumigashira}
\affiliation{Institute of Multidisciplinary Research for Advanced Materials (IMRAM), Tohoku University, Sendai, 980–8577, Japan}

\author{Keiichiro Imura}
\affiliation{Department of Physics, Nagoya University, Nagoya 464-8602, Japan}

\author{Hiroyuki S. Suzuki}
\affiliation{Institute for Solid State Physics, The University of Tokyo, Kashiwa 277-8581, Japan}

\author{Noriaki K. Sato}
\affiliation{Department of Physics, Nagoya University, Nagoya 464-8602, Japan}
\affiliation{Center for General Education, Aichi Institute of Technology, Toyota 470-0392, Japan}

\author{Shin-ichi Kimura}
\email{kimura.shin-ichi.fbs@osaka-u.ac.jp}
\affiliation{Graduate School of Frontier Biosciences, Osaka University, Suita 565-0871, Japan}
\affiliation{Department of Physics, Graduate School of Science, Osaka University, Toyonaka 560-0043, Japan}
\affiliation{Institute for Molecular Science, Okazaki 444-8585, Japan}

\date{\today}

\begin{abstract} 
The electronic structure changes of SmS surfaces under potassium (K) doping are elucidated using synchrotron-based core-level photoelectron spectroscopy and angle-resolved photoelectron spectroscopy (ARPES).
The Sm core-level and ARPES spectra indicate that the Sm mean valence of the surface increased from the nearly divalent to trivalent states, with increasing K deposition.
Carrier-induced valence transition (CIVT) from Sm$^{2+}$ to Sm$^{3+}$ exhibits a behavior opposite to that under conventional electron doping.
Excess electrons are trapped by isolated excitons, which is inconsistent with the phase transition from the black insulator with Sm$^{2+}$ to the gold metal with Sm$^{3+}$ under pressure.
This CIVT helps to clarify the pressure-induced black-to-golden phase transition in this material, which originates from the Mott transition of excitons.
\end{abstract}

\maketitle
At the boundary between localized and itinerant electron systems in solids, fluctuations in the degrees of freedom of the electrons result in a variety of physical properties, such as unconventional superconductivity and giant magnetoresistance.
Among such systems, in the so-called heavy-fermion systems containing rare-earth ions, the physical properties change from non-magnetic heavy quasiparticles originating from the Kondo effect appears to a magnetic ordering due to the Ruderman–Kittel–Kasuya–Yosida (RKKY) interactions, depending on the strength of the hybridization between the conduction electrons and 4$f$ electrons, namely $c$-$f$ hybridization \cite{Gegenwart, Pfleiderer}.
Phase transitions are often accompanied by valence transitions \cite{Miyake}.
Changes in the hybridization strength and the resulting variations in the physical properties caused by the valence transition have attracted considerable attention.

Samarium mono-sulfide (SmS) is a heavy-fermion system, whose physical properties vary drastically when the valence of the Sm ions changes \cite{Jayaraman70}.
At ambient pressure, SmS acts as a semiconductor with an indirect bandgap of about 0.1 eV \cite{kimura08} and appears black \cite{Matsubayashi071}.
The electrical properties change to metallic above a critical pressure of about  0.65 GPa \cite{Keller79, Raymond02, Annese06, Mizuno08, Imura20}.
Metallic (golden phase) SmS is predicted to possess a topologically nontrivial surface state \cite{Kang19}, similar to the topological surface state of Kondo insulators such as SmB$_{6}$ \cite{Ohtsubo19} and YbB$_{12}$ \cite{Hagiwara16}.
Although this pressure-induced black insulator-to-golden metal phase transition (BGT) was discovered more than 50 years ago \cite{Jayaraman70}, the origin of this phase transition still remains under debate \cite{watanabe21}.
Recently, a phase transition like the pressure-induced one has been reported due to two different types of perturbations: light irradiation and current injection \cite{Kitagawa03, Ando20}.
These perturbations are commonly expected to increase carriers, namely carrier-induced valence transition (CIVT), but the detailed electronic states after carrier doping have not been clarified yet.

Alkali metal adsorption on crystal surfaces is a widely used method to investigate the effects of carrier injection into materials \cite{Aruga, Diehl}.
Owing to the electron doping from the alkali metal to the sample, the sample surface becomes negatively doped, sometimes a band structure rigidly shifts to the higher binding energy side.
The doping concentration can be precisely manipulated by controlling the total amount of deposition.
This technique is occasionally combined with angle-resolved photoelectron spectroscopy (ARPES) to observe the unoccupied electronic states \cite{Zhu11, Zhang14}, and is also employed to observe the negative electron affinity on semiconductor surfaces, owing to the formation of the surface electric double layer \cite{Levine73}.
Therefore, through alkali metal adsorption, a well-defined, carrier-doped SmS surface can be realized.
Such a surface would serve as a suitable candidate for investigating the carrier-injection effect.

\begin{figure*}
\includegraphics[width=150mm]{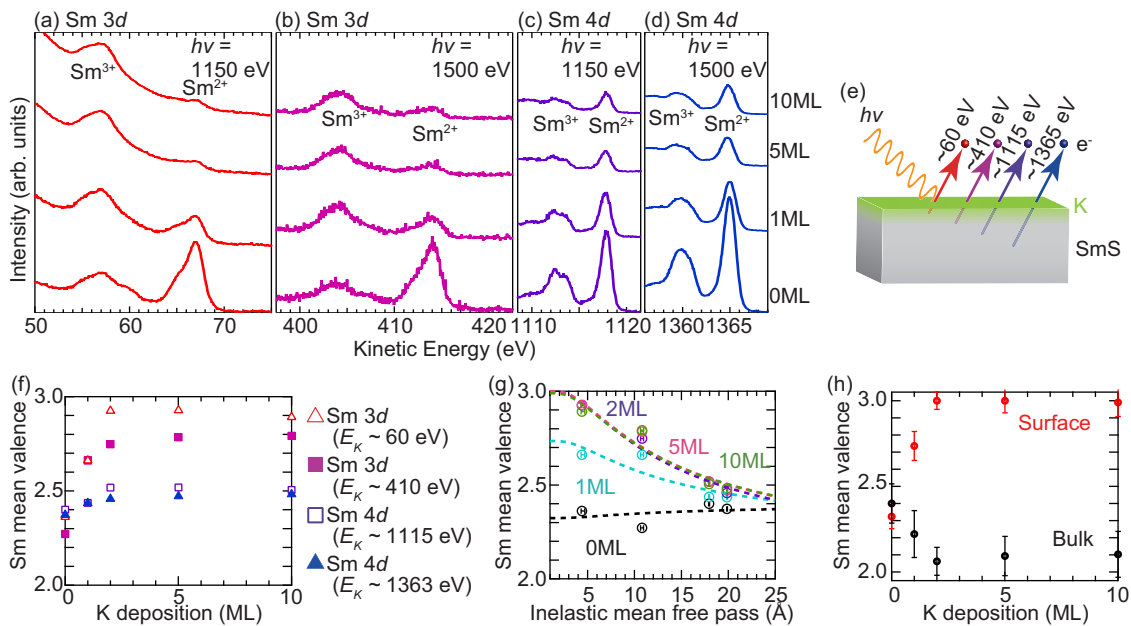}
\caption{\label{figure 1}
(a,b) Sm 3$d$ and (c,d) Sm 4$d$ core-level spectra of SmS single crystals as a function of the amount of K deposition, with excitation photon energies at (a,c) 1150 eV and (b,d) 1500 eV.
(e) Schematic of the photoemission detection depth for different kinetic energies.
(f) Sm mean valence depending on K deposition, evaluated from the peak area ratio between the Sm$^{3+}$ and Sm$^{2+}$ multiplets after subtracting the Shirley-type background.
(g) Sm mean valence as a function of the mean free path of photoelectrons.
Dashed lines indicate the fitting curves expected with a two-layer model, shown in supplementary material S3.
Horizontal bars are the different mean free paths for the highest and lowest kinetic energies of the Sm multiplets.
(h) Surface and bulk Sm mean valences as a function of the amount of K deposition, evaluated with the two-layer model.
}
\end{figure*}
In this Letter, we have studied the change of the electronic structure by potassium (K) doping on a SmS surface by synchrotron-based ARPES and core-level photoelectron spectroscopies.
With increasing the amount of the K deposition, the Sm mean valence at the surface was increased from nearly divalent to trivalent, which is the opposite behavior observed in many materials, i.e., the mean valence normally decreases due to electron donation from the adsorbed alkali metal to the surface.
This CIVT plays an important role in terms of clarifying the pressure-induced BGT in this material.

\begin{figure}
\includegraphics[width=80mm]{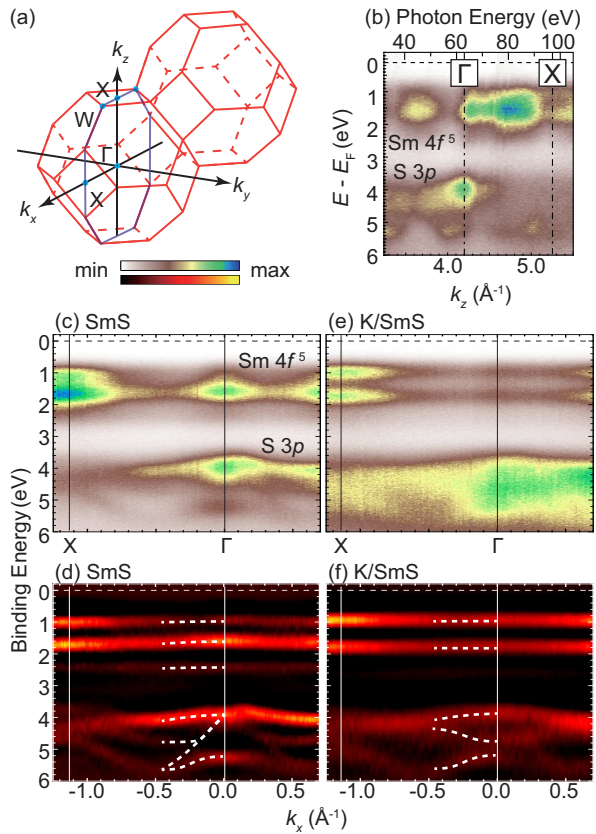}
\caption{\label{figure 2}
(a) Three-dimensional bulk Brillouin zone of SmS.
(b) Photon-energy dependence of valence band spectra at normal emission ($k_x$ = $k_y$ = 0 \AA$^{-1}$).
Intensities of the photoelectron are normalized by maximum counts of the valence band spectra at each excitation energy in the measured energy range (binding energy = 0 $\sim$ 6.0 eV).
Inner potential was set to 14.1 eV, based on the periodicity of the S 3$p$ band around 4.5 eV.
(c) ARPES intensity plot and (d) second-derivative of the ARPES intensity of the pristine-SmS surface along $\Gamma$--$\rm X$, with 60 eV photons ($k_z$ $\sim$ 0 \AA$^{-1}$), respectively.
(e,f) Same as (c,d) but for the K-adsorbed SmS ($\rm K/SmS$) surface.
White dashed lines serve as guides for the band dispersion.
}
\end{figure}

High-quality single-crystalline SmS was grown using the vertical Bridgman method in a high-frequency induction furnace \cite{Matsubayashi071}.
A clean SmS surface was obtained via cleaving \textit{in-situ} in an ultra-high vacuum chamber at room temperature (RT).
K atoms were evaporated with a well-degassed alkali metal dispenser (SAES-Getters) at RT.
The adsorption of K atoms on the samples was examined using core-level photoelectron spectroscopy, as shown in Fig. S1 \cite{SM}.
The thickness of K atoms was estimated by using quartz microbalance and calibrated by the intensity of the K 3\textit{p} and Sm 4\textit{p} core levels.
In this work, one monolayer (ML) of the evaporated K atoms was defined as the atomic density of bulk SmS.

ARPES and core-level photoelectron spectroscopy measurements were performed at BL5U of UVSOR-III and BL-2A MUSASHI of Photon Factory, with photon energies ranging from 35 to 1500 eV with $p$-polarization.
The energy resolutions of the ARPES and core-level photoelectron spectroscopy were set to $\sim$30 and $\sim$100 meV, respectively.
The energy resolution and position of the Fermi level were calibrated using the Fermi edge of the evaporated Au film.
All the measurements were performed at RT.

Figures 1(a)-1(d) show the core-level spectra of Sm 3$d$ and 4$d$ as a function of the K deposition thickness.
From Fig. 1(a), the Sm trivalent (Sm$^{3+}$) and divalent (Sm$^{2+}$) multiplet peaks are observed at kinetic energies of  $\sim$58 and $\sim$68 eV, respectively.
For 0-ML deposition (pristine-SmS), the intensity of the Sm$^{2+}$ peak was higher than that of the Sm$^{3+}$ peak, suggesting that the Sm ions were nearly divalent.
With increasing K deposition thickness, the Sm$^{3+}$ component gradually became dominant.
After the deposition of more than 5 ML, the core-level spectral shape remained unchanged, suggesting the saturation of K adsorption.
Since the substrate temperature was kept at RT through the whole experiment, the adsorption would be limited to a few monolayers, similar to that of other alkali metal adsorbed on semiconductor systems such as Cs/Si(001) surface \cite{Holtom77}.

As shown in Figs. 1(b)-1(d), the Sm$^{3+}$ component became dominant with an increase in the K deposition thickness, similar to the results shown in Fig. 1(a). However, the rate of increase decreased as the kinetic energy of the photoelectrons increased.
These results can be explained by the kinetic energy dependence of the effective traveling length of photoelectrons, as illustrated in Fig. 1(e).
In the present case, the Sm 3$d$ and 4$d$ core levels excited by 1150 eV (Fig. 1(a)) and 1500 eV (Fig. 1(d)) photons are expected to be the most surface- and bulk-sensitive, respectively.

To obtain further information regarding the change in the Sm mean valence due to the K adsorption, the mean valence was evaluated from the peak area ratio between the Sm$^{2+}$ and Sm$^{3+}$ components in the core-level spectrum after subtracting the Shirley-type background, as shown in Fig. 1(f).
With reference to the typical relationship between the kinetic energy and the inelastic mean free path of photoelectrons \cite{IMFP}, the mean valences as a function of the photoelectron mean free path were obtained, as indicated by the marks in Fig. 1(g).
The best fit and the obtained surface and bulk mean valences, determined via fitting with the formula in Supplementary Note 3, are presented as dashed lines in Fig. 1(g).
Using these fitting parameters, the mean valences on the surface and in the bulk as a function of the K deposition thickness can be evaluated, as shown in Fig. 1(h).
In pristine-SmS, the mean valence at the surface was slightly closer to the divalent state than that of the bulk, as is generally observed in valence fluctuation materials \cite{hagiwara17}, because of the lattice expansion at the surface.
The mean valence of the surface component increased with the K deposition thickness below 2 ML.
This behavior, however, is not typical because alkali metal adsorption is usually effective for electron doping.
The maximum value of the surface mean valence was approximately 3.0, which is almost identical to that of the golden phase of SmS under high pressures.
By contrast, in the bulk region, the mean value changed slightly from $2.4 \pm 0.1$ to $2.1 \pm 0.15$ with the increasing K content.
These results suggest that the mean valence of the Sm ions on the SmS surface can be easily increased via electron doping, even though the bulk valence is conserved.
It should be noted that the absorption spectrum at the Sm $L$ edge suggests that the bulk mean valence of Sm at ambient pressure is approximately divalent \cite{Imura20}, which is inconsistent with our data. This discrepancy can be due to different probing depth in different methods.

\begin{figure}
\includegraphics[width=80mm]{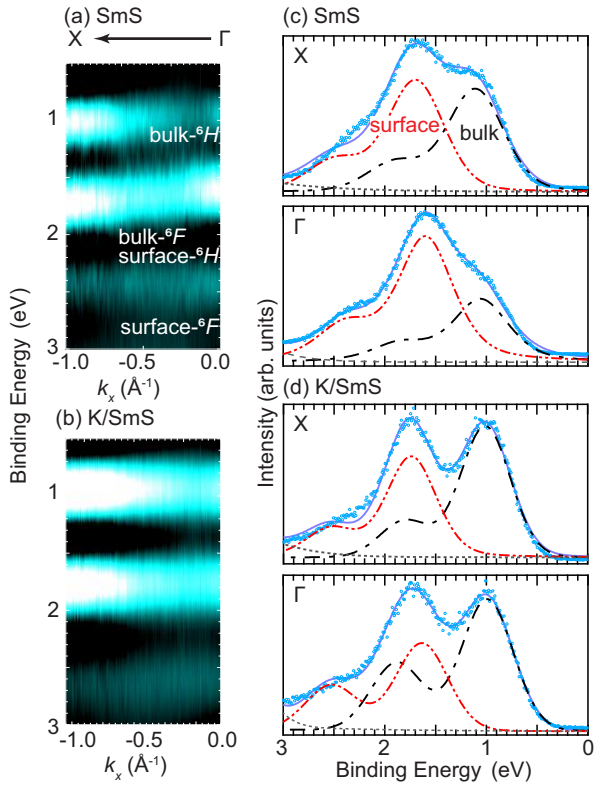}
\caption{\label{figure 3}
(a-b) Second-derivative of ARPES intensity maps near the Fermi level along $\Gamma$--$\rm X$, with 60 eV photons of (pristine-) SmS and K/SmS.
$^6$\textit{H} and $^6$\textit{F} indicate the multiplet structures of the Sm 4$f$$^5$ final state after photo-excitation.
(c) Energy distribution curves (EDCs) at the $\Gamma$ point (lower panel) and the $\rm X$ point (upper panel) of the SmS surface with the fitting curves.
Dots and lines represent the raw data and fitted spectra (dash-dotted line: bulk component, dash-double-dotted line: surface component, and solid line: sum of the bulk and surface components).
Each component is fitted by the Voigt function after subtracting the Shirley-type background  (dashed line).
(d) Same as (c) but for the K/SmS.
}
\end{figure}

\begin{figure}
\includegraphics[width=80mm]{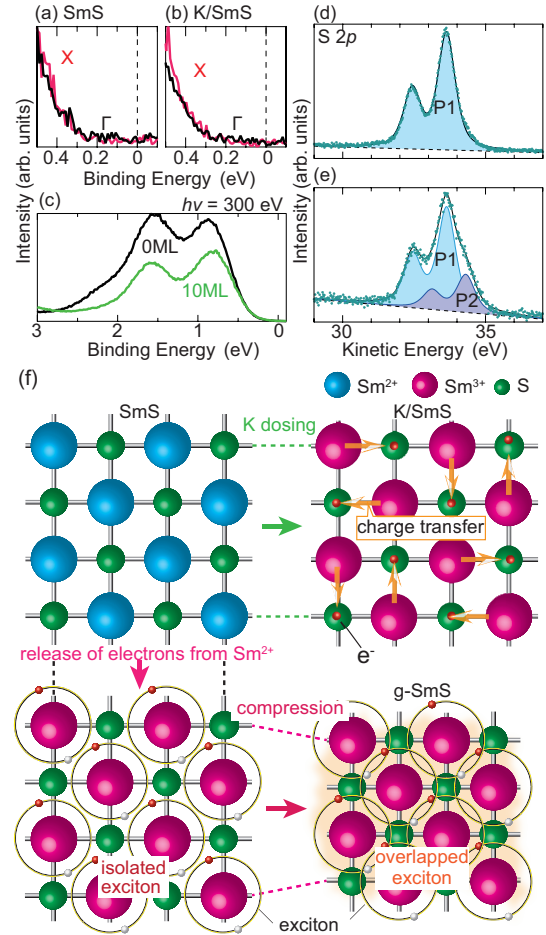}
\caption{\label{figure 4}
(a) EDCs at the $\Gamma$ point (black line) and $X$ point (green line) of pristine-SmS with 200 eV photons.
(b) Same as (a) but for K/SmS.
(c) Angle-integrated valence band spectra of SmS and K/SmS with 300-eV photons.
S 2$p$ core-level spectra of (d) pristine-SmS and (e) K/SmS with 200-eV photons.
Dashed lines indicate the Shirley-type background.
Blue dots and black lines represent the raw data and fitted curve using the Voigt function and background.
(f) Schematic of the phase transition of SmS.
Top left and top right images depict the real space image of pristine-SmS and K/SmS, respectively.
The large (small) sphere indicates the Sm (S) ion.
The bottom left figure is an expected model of electron-doped SmS, while retaining the same lattice constant as pristine-SmS, in contrast to the golden phase of SmS, by applying pressure, as shown in the bottom right figure.
Yellow circles depict the electron-hole pairs (excitons).
}
\end{figure}

The change of the mean valence is expected to be strongly influenced in the valence band ARPES spectra.
The ARPES  spectra of K-adsorbed SmS (K/SmS), as compared with those of pristine-SmS, are shown in Fig. 2.
The bulk Brillouin zone (BZ) of SmS is shown in Fig. 2(a).
To identify the $\Gamma$ point along the surface normal ($k_z$) wavenumber, the excitation photon-energy dependence of the valence band spectrum at $k_x$ = $k_y$ = 0 \AA$^{-1}$ was measured, as shown in Fig. 2(b).
The multiplet structure of the flat Sm$^{2+}$ 4\textit{f}$^{5}$ final states and dispersive S 3\textit{p} bands were observed near $\sim$1.5 and $\sim$4.5 eV, respectively, \cite{ito, SmMilti}.
By setting the inner potential to 14.1 eV from the folded period of the S 3\textit{p} band, which is identical to that previously evaluated \cite{Kaneko}, $\textit{h}\nu$$\sim$60 eV and $\sim$95 eV correspond to the $\Gamma$ and $\rm X$ points along the $k_z$ direction, respectively.
Figure 2(c) shows the ARPES intensity plot of the pristine-SmS along $\Gamma$--$\rm X$, as obtained with 60-eV photons.
As well as the $k_z$ direction shown in Fig. 2(b), flat Sm$^{2+}$ 4\textit{f}$^{5}$  multiplet and dispersive S 3\textit{p} band were observed.
The second-derivative ARPES image, highlighting the shape of the band dispersion, is shown in Fig. 2(d).
The 4\textit{f}$^{5}$ multiplets can be decomposed into three components that are weakly dispersed from the $\Gamma$ point to the $\rm X$ point.
At a binding energy of $\sim$4 eV, the three dispersive bands degenerate near the $\Gamma$ point.
Figures 2(e) and 2(f) present ARPES images and a second-derivative plot, respectively, of approximately 0.4 ML for the K/SmS surface.
On many alkali metal-adsorbed surfaces, electron doping causes a band shift toward the side with high binding energy; however, no such change was observed in this case.
This result suggests that the change of the electronic structure of the K/SmS surface cannot be explained by a simple rigid band shift model due to electron doping.

To obtain further insights into the deformation of the band structure owing to K adsorption, magnified images of the second-derivative ARPES for the 4\textit{f} bands are shown in Figs. 3(a) and 3(b).
In pristine-SmS, the three branches located at 1, 1.7, and 2.5 eV in the 4\textit{f} bands observed in ARPES are assigned to the multiplets of bulk-$^6$$H$, sum of bulk-$^6$$F$ and surface-$^6$$H$, and surface-$^6$$F$, respectively.
Figures 3(c) (3(d)) show the energy distribution curves at the $\Gamma$ and $\rm X$ points of the pristine-SmS and K/SmS surfaces, respectively.
To clarify the contributions of the surface and bulk components, the peaks were separated by Gaussian fittings.
After K deposition, the surface components were suppressed; this is qualitatively consistent with the behavior of the core-level peaks in Fig. 1.

The core-level photoelectron spectra and ARPES results commonly suggest that CIVT from Sm$^{2+}$ to Sm$^{3+}$ occurs on the SmS surface; however, this cannot be explained from the perspective of a simple rigid band picture.
We now discuss the role of the excess electrons generated by the change from Sm$^{2+}$ to Sm$^{3+}$.
In the pressure-induced golden phase attained via BGT, the excess electrons became conduction electrons.
However, the overall electronic state near the Fermi level retains a semiconducting nature, without any change caused by the K adsorption, as shown in Figs. 4(a) and 4(b).
In other words, the surface electronic state did not become metallic under K-doping.
Also, in the bulk region, the semiconducting feature is kept as shown in Fig. 4(c).
On the other hand, as shown in Figs. 4(d) and 4(e), the S~$2p$ core-level peak is a single component in pristine-SmS, with only two peaks separated by spin-orbit interaction (SOI), while in K/SmS, a new SOI-separated peak appears at the low energy side of the double-peak observed in pristine-SmS.
The S ion in pristine-SmS had a single valence number (S$^{2-}$), whereas that in K/SmS featured two types of valences, suggesting the appearance of S$^{-2-\delta}$ ions.
Above these results, it is expected that the free electrons from K atoms are transferred to SmS topmost surface layer and they are preferentially bounded to the S atoms, which is the origin of S$^{-2-\delta}$ valence state (P2 in Fig. 4(e)).
Then, the valence of the Sm atoms is increased to conserve the charge-neutral condition between Sm and S.
Considering that K/SmS remained in the semiconducting state and the Sm valence of the surface was trivalent, the excess electrons generated by the appearance of Sm$^{3+}$ can be regarded as being trapped at the S site.

These results were compared with the electronic state of the gold phase under pressure.
When pressure is applied, the lattice parameter decreases by approximately 5~\% owing to the decrease in the ionic radius of Sm by the BGT \cite{Imura20}, and a carrier density similar to that of a monovalent metal is developed.
(This is the origin of the golden color.)
However, in the case of K/SmS, the bulk electronic state remained unchanged, suggesting that the bulk lattice constant was almost unchanged.
Because the surface lattice constant does not differ significantly from the bulk lattice constant, it was considered to be almost unchanged.
Under these conditions, conduction electrons do not appear, even if the Sm ions on the surface become trivalent, suggesting that the conduction electrons in the pressure-induced golden phase originate from the decrease in the lattice constant.
Thus, the origin of these conduction electrons in the golden phase is attributed to the decrease in the lattice parameter.

These results are consistent with those for the insulator-to-metal transition caused by the Mott transition of excitons \cite{Guerci19}.
If the critical distance of the Mott transition of an exciton represents the lattice constant at the critical pressure of the BGT, then the BEC--BCS phase transition of the exciton occurs when the exciton size exceeds the critical lattice constant, as shown in Fig. 4(f).
In the case of a change in the surface valence alone, there is no variations in the lattice constant; hence, even if electrons are released from the Sm ions, they are trapped by the isolated excitons.
Therefore, the phase remains in the BEC state, which is consistent with the fact that no metal-insulator transition appears.
This result strongly suggests that the CIVT is not directly linked with the insulator-metal transition and may provide useful information regarding the origin of the pressure-induced BGT.

In conclusion, we have studied the electronic structure on K adsorbed SmS surface by using synchrotron-based core-level photoelectron spectroscopies and ARPES.
With an increase in the amount of K deposition, the mean valence of Sm at the surface increased from the nearly divalent to trivalent states, without a rigid band shift.
This result suggests that the injected carriers induce the valence transition of the Sm ions from Sm$^{2+}$ to Sm$^{3+}$; this is opposite to the charge transition observed under trivial electron doping, without an insulator-metal transition in the pressure-induced phase transition.
The lack of metallization is considered to originate from the carriers being trapped by isolated excitons.

We thank T. Ito and H. Watanabe for the helpful discussions.
The ARPES and core-level PES measurements were performed under the UVSOR proposal (21-666, 21-855) and Photon Factory proposal (2019G514).
This study was supported by JSPS KAKENHI (Grant No. JP20K03859, JP19H01830, and JP20H04453) and the Murata Science Foundation.

\subsection{Supplementary material for Surface valence transition in SmS by alkali metal adsorption}

\subsection{S1. Wide range valence band dispersion}
In pristine-SmS, the Sm 4$f$ multiplet peaks are observed only in the measured spectral range.
After K deposition, the K 3$p$ and weak K 3$s$ states were visible; this serves as evidence of the K atoms adsorbed on the SmS surface.
Multiplet peaks of the Sm trivalent final states were also observed around the kinetic energy of 185 eV.

\subsection{S2. Sm 4$p$$_{3/2}$ and K 2$p$ core-level spectra of the 20-ML K-adsorbed SmS}
Figure S2 shows the relative photoelectron intensities of (a) Sm 4\textit{p}$_{3/2}$ and (b) K 2\textit{p} core-levels of the 20-ML K-adsorbed SmS surface taken with 1486.6-eV photons.
Note that a cross-section of the photoemission process between Sm 4\textit{p}$_{3/2}$ and K 2\textit{p}$_{3/2}$ is about 1.4 times larger in Sm 4\textit{p} \cite{Yeh}.
The photoelectron intensities in both the core levels of Sm and K are comparable even with 20 ML of K deposited, which suggests that K adsorption is limited to a few layers on the surface and diffusion into the bulk is little.

\subsection{S3. Calculation of the depth dependence of the Sm mean valence  (Two-layer model)}
The Sm mean valence {\it v} can be described by
\begin{equation}
v = 2 + \frac{n_{3+}}{n_{2+}+n_{3+}} = 2 + {N_{3+}},
\end{equation}
where $n_{2+}$ and $n_{3+}$ are the photoelectron peak area intensities of the Sm$^{2+}$ and Sm$^{3+}$
components, respectively, in the Sm core-level spectra, and $N_{3+}$ is the normalized intensity of the Sm$^{3+}$ component.
The Sm mean valence, depending on the inelastic mean free path $v_{I}$(l), can be obtained as follows:
\begin{equation}
v_{I}(l)= 2 + \int^{\infty}_{0} \frac{{N_{3+}(x)}e^{-{\frac{x}{l}}}}{l} dx,
\end{equation}
where {\it x} and {\it l} are the thickness of the topmost surface layer and the inelastic mean free path of the photoelectrons, respectively.
To simplify the evaluation of the depth dependence of the Sm mean valence, $v_{I}$(l) is decomposed into two terms, assumed as two layers of a surface layer and a bulk layer (two-layer model), yielding the following equation:
\begin{equation}
v_{I}(l)\simeq 2 + \int^{d}_{0} \frac{{N_{3+}^{surface}}e^{-{\frac{x}{l}}}}{l} dx +\int^{\infty}_{d} \frac{{N_{3+}^{bulk}}e^{-{\frac{x}{l}}}}{l} dx,
\end{equation}
where {\it d} is the depth of the surface region, and ${N_{3+}^{surface}}$ and ${N_{3+}^{bulk}}$ are the normalized Sm$^{3+}$ intensities at the surface and bulk, respectively.
In this study, {\it d} was assumed to be 12 \AA, which is consistent with a thickness of twice the lattice constant of SmS.
By fitting the relationship between the inelastic mean free path and the Sm mean valence, as shown in Fig. 1(g), using Eq. (3), the depth-dependent Sm mean valences were obtained, as shown in Fig. 1(h).

\begin{figure}
\includegraphics[width=80mm]{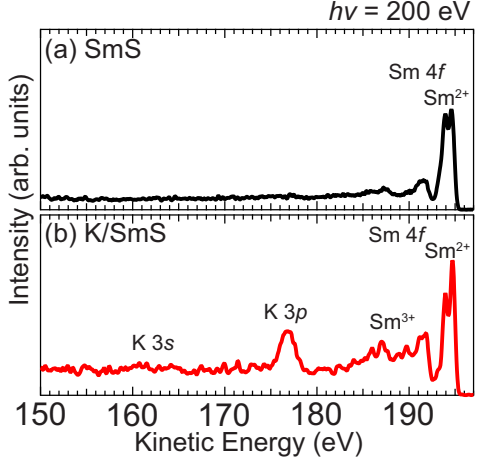}
\caption{\label{figure S1}
Wide range valence band spectra of (a) SmS and (b) K/SmS excited by 200 eV photons.
}
\end{figure}

\begin{figure}
\includegraphics[width=80mm]{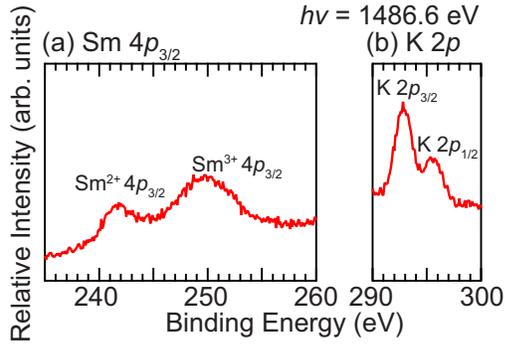}
\caption{\label{figure S2}
Core-level spectra of (a) Sm 4\textit{p}$_{3/2}$ and (b) K 2\textit{p} of 20-ML deposited K/SmS excited by 1486.6 eV photons.
}
\end{figure}

\end{document}